\documentclass[preprint]{vgtc}               

\ifpdf
  \pdfoutput=1\relax                   
  \usepackage{graphicx}                
  \DeclareGraphicsExtensions{.pdf,.png,.jpg,.jpeg} 
\else
  \usepackage{graphicx}                
  \DeclareGraphicsExtensions{.eps}     
\fi%

\graphicspath{{figures/}{pictures/}{images/}{./}} 

\usepackage{microtype}                 
\PassOptionsToPackage{warn}{textcomp}  
\usepackage{textcomp}                  
\usepackage{mathptmx}                  
\usepackage{times}                     
\usepackage{cite}     
\usepackage{tabu} 
\usepackage{booktabs} 
\usepackage{changes}

\onlineid{0}

\vgtccategory{Research}

\vgtcinsertpkg

\title{Facilitating Conversational Interaction\\ in Natural Language Interfaces for Visualization}

\author{Rishab Mitra$^\pi$\thanks{e-mail: rmitra34@gatech.edu} \and Arpit Narechania$^\pi$\thanks{e-mail: arpitnarechania4@gatech.edu} \and Alex Endert\thanks{e-mail: endert@gatech.edu} \and John Stasko\thanks{e-mail: stasko@cc.gatech.edu\newline\indent$^\pi$authors contributed equally}} \affiliation{\scriptsize Georgia Institute of Technology}

\abstract{
Natural language (NL) toolkits enable visualization developers, who may not have a background in natural language processing (NLP), to create natural language interfaces (NLIs) for end-users to flexibly specify and interact with visualizations.
However, these toolkits currently only support one-off utterances, with minimal capability to facilitate a multi-turn dialog between the user and the system.
Developing NLIs with such conversational interaction capabilities remains a challenging task, requiring implementations of low-level NLP techniques to process a new query as an intent to follow-up on an older query.
We extend an existing Python-based toolkit, \app, that processes an NL query about a tabular dataset and returns an analytic specification containing data attributes, analytic tasks, and relevant visualizations, modeled as a JSON object. Specifically, \app now enables developers to facilitate multiple simultaneous conversations about a dataset and resolve associated ambiguities, augmenting new conversational information into the output JSON object.
We demonstrate these capabilities through three examples: (1) an NLI to learn aspects of the Vega-Lite grammar, (2) a mind mapping application to create free-flowing conversations, and (3) a chatbot to answer questions and resolve ambiguities.

} 

\CCScatlist{
  \CCScatTwelve{Human-centered computing}{Visu\-al\-iza\-tion}{Visu\-al\-iza\-tion systems and tools}{Visu\-al\-iza\-tion toolkits};
  \CCScatTwelve{Human-centered computing}{Human computer interaction (HCI)}{Interaction techniques}{Text input}
}

\usepackage{microtype}                 
\PassOptionsToPackage{warn}{textcomp}  
\usepackage{textcomp}                  
\usepackage{mathptmx}                  
\usepackage{times}                     
\usepackage{cite}                      
\usepackage{tabu}                      
\usepackage{booktabs}                  
\usepackage[font=sf]{caption}
\usepackage[normalem]{ulem}
\usepackage{enumitem}
\usepackage{calc}
\usepackage{textcomp}
\usepackage{graphicx}
\usepackage{soul}
\usepackage{xcolor}
\usepackage{tcolorbox}
\usepackage[frozencache,cachedir=.]{minted}

\usepackage{fontawesome}
\usepackage{multirow}
\usepackage{hyperref}
\usepackage{xspace}

\newcommand{\add}[1]{#1}
\newcommand{\remove}[1]{\textcolor{pink}{}} 

\definecolor{orange}{RGB}{237,125,49}
\definecolor{purple}{RGB}{112,47,160}
\definecolor{pink}{RGB}{245,114,255}
\definecolor{black}{RGB}{0,0,0}

\definecolor{customLightGreen}{HTML}{D9EAD3}
\definecolor{customCodeBlue}{HTML}{214A87}
\definecolor{customGreen}{HTML}{54AA54}
\definecolor{customCodeOrange}{HTML}{C55A11}
\definecolor{darkteal}{HTML}{004B53}
\definecolor{darkred}{HTML}{A50021}
\definecolor{urlblue}{HTML}{067DE9}

\definecolor{explicitBlack}{HTML}{404040}
\definecolor{ambiguousGray}{HTML}{DEDEDE}

\definecolor{explicitBlue}{HTML}{0070C0}
\definecolor{ambiguousBlue}{HTML}{CCECFF}
\definecolor{implicitBlue}{HTML}{002060}

\definecolor{key0Color}{HTML}{5FaD56}
\definecolor{key1Color}{HTML}{ca7611}
\definecolor{key2Color}{HTML}{ca7611}
\definecolor{inputParamColor}{HTML}{BA2121}
\definecolor{inputParamColorBool}{HTML}{008001}
\definecolor{valueColor}{HTML}{4F9907}

\newcommand{\inputParamValue}[1]{\textcolor{inputParamColor}{#1}}
\newcommand{\variable}[1]{\textcolor{customCodeBlue}{\texttt{\textbf{#1}}}}
\newcommand{\function}[1]{\texttt{\textbf{#1}}}

\newcommand{\val}[1]{\textcolor{valueColor}{#1}}

\newcommand{\app}{NL4DV\xspace}

\begin{document}


\firstsection{Introduction and Background}

\maketitle

\label{section:introduction}
Natural language interfaces (NLIs) for databases~\cite{blunschi2012soda, li2014nalir, pasupat2015compositional, zhong2017seq2sql, wang2018robust, he2019x, herzig2020tapas, narechania2021diy} and visualizations~\cite{mspowerbi, tableauaskdata, amazonquicksight, sun2010articulate, gao2015datatone, setlur2016eviza, hoque2017applying, srinivasan2018orko, yu2019flowsense, srinivasan2020inchorus, setlur2019inferencing, srinivasan2020interweaving, kumar2016towards, kassel2018valletto, kim2020answering, narechania2020nl4dv} have shown great promise, democratizing access to data through the querying power and expressivity of natural language (NL). 
Given a dataset (e.g., movies), an NLI for visualization receives an NL query (e.g., \emph{``Show the distribution of budget''}) as input, extracts data attributes (\emph{Production Budget}) and analytic tasks (\emph{Distribution}), and recommends one or more relevant visualizations (\emph{Histogram}). 
Many of these NLIs also help resolve ambiguities that may occur during query interpretation. For example, DataTone~\cite{gao2015datatone} presents ambiguities through interactive GUI-based widgets (e.g., dropdowns) for disambiguation.
Implementing such NLIs, however, requires experience with natural language processing (NLP) techniques and toolkits (e.g., NLTK~\cite{loper2002nltk}, spaCy~\cite{honnibal2017spacy}) as well as GUI and visualization design tools (e.g., D3.js~\cite{bostock2011d3}, Vega-Lite~\cite{satyanarayan2016vega}), making it challenging for developers without the necessary skillset.

Recently, NL toolkits~\cite{narechania2020nl4dv, fu2020quda, liu2021advisor, luo2021natural} have enabled visualization developers, who may not have a background in NLP, to create new visualization NLIs or incorporate NL input within their existing systems. 
For example, given a tabular dataset and an NL query about the dataset, \app generates an analytic specification comprising data attributes, analytic tasks (based on~\cite{amar2005low}), and visualizations (as Vega-Lite specifications~\cite{satyanarayan2016vega}) modeled as a JSON object.
\remove{Similarly, ncNet~\cite{luo2021natural}, a Transformer based sequence-to-sequence (seq2seq) model trained using nvBench~\cite{luo2021synthesizing}, adopts a self-attention mechanism to generate a rich representation of the input dataset to recommend visualizations.}
However, these toolkits currently support one-off utterances (singleton queries) only, with minimal capability to facilitate a multi-turn dialog between the end-user and the system, e.g., by following-up on a previous query.
Because of this, end-users would have to specify longer NL queries (e.g., \emph{``Show the relationship between budget and rating for Action and Adventure movies that grossed over 100M'')} to accomplish more complex tasks. These types of queries may also have a greater chance of failing (e.g., attribute detection can fail; filter operators may be incorrect), eventually warranting several paraphrasing attempts. 
We believe specifying multiple short queries in a natural sequence can enable end-users to incrementally accomplish a complex task, fix minor errors, and also make debugging easier, as in~\cite{setlur2016eviza,hoque2017applying,srinivasan2020interweaving,srinivasan2021snowy,amazonalexa,googlehome}.
This is called \emph{conversational interaction} -- ``face-to-face or technology-mediated forms of interaction that use language, encompassing a wide range of different types of talk''~\cite{haugh2012conversational}.

Developing NLIs with such conversational interaction capabilities remains a challenging task, however, requiring implementations of low-level NLP techniques to process a new query as an intent to follow-up on an older query, e.g., replacing an existing attribute with a new one.
To the best of our knowledge, no NL toolkit facilitates conversational interaction, yet.
Hence, in this work, we extend a Python-based toolkit, \app~\cite{narechania2020nl4dv}, in order to enable visualization developers to facilitate multiple simultaneous conversations (through manual specification as well as automatic detection of intents to follow-up) and resolve associated ambiguities through an easy-to-use application programming interface (API). As a result, \app also augments additional conversational information into the output JSON.
We demonstrate these capabilities through three examples: (1) an NLI to learn aspects of \remove{the Vega-Lite grammar}\add{Vega-Lite -- an implementation of a grammar for interactive graphics~\cite{satyanarayan2016vega}}, (2) a mind mapping application to create free-flowing conversations about a dataset, and (3) a chatbot to answer questions and resolve ambiguities in collaboration with the enduser. To support development of future systems, we open-source \app and the described applications at \textbf{\url{https://nl4dv.github.io/nl4dv/}}.






\section{Conversational Interaction with \app}
\label{section:app}
\begin{listing}[t!]

\begin{minted}
    [
    baselinestretch=1,
    fontsize=\small,
    xleftmargin=15pt,
    linenos,
    breaklines,
    escapeinside=||,
    % ,style=tango
    ]
    {python}
from nl4dv import NL4DV
nl4dv_instance = NL4DV(data_url="housing.csv")
resp_1 = nl4dv_instance.analyze_query("Show average prices for different home types over the years.")
print(resp_1)
# a new dialogId and a queryId get created.
\end{minted}

\vspace{-1em}

\begin{minipage}[!b]{4cm}
    \begin{minted}[fontsize=\small,style=tango]{json}
    {
        "dialogId": "0",
        "queryId": "0", 
        ...
    }
    \end{minted}
\end{minipage}\begin{minipage}[!b]{3.1cm}
    \centering
    \includegraphics[width=3.1cm]{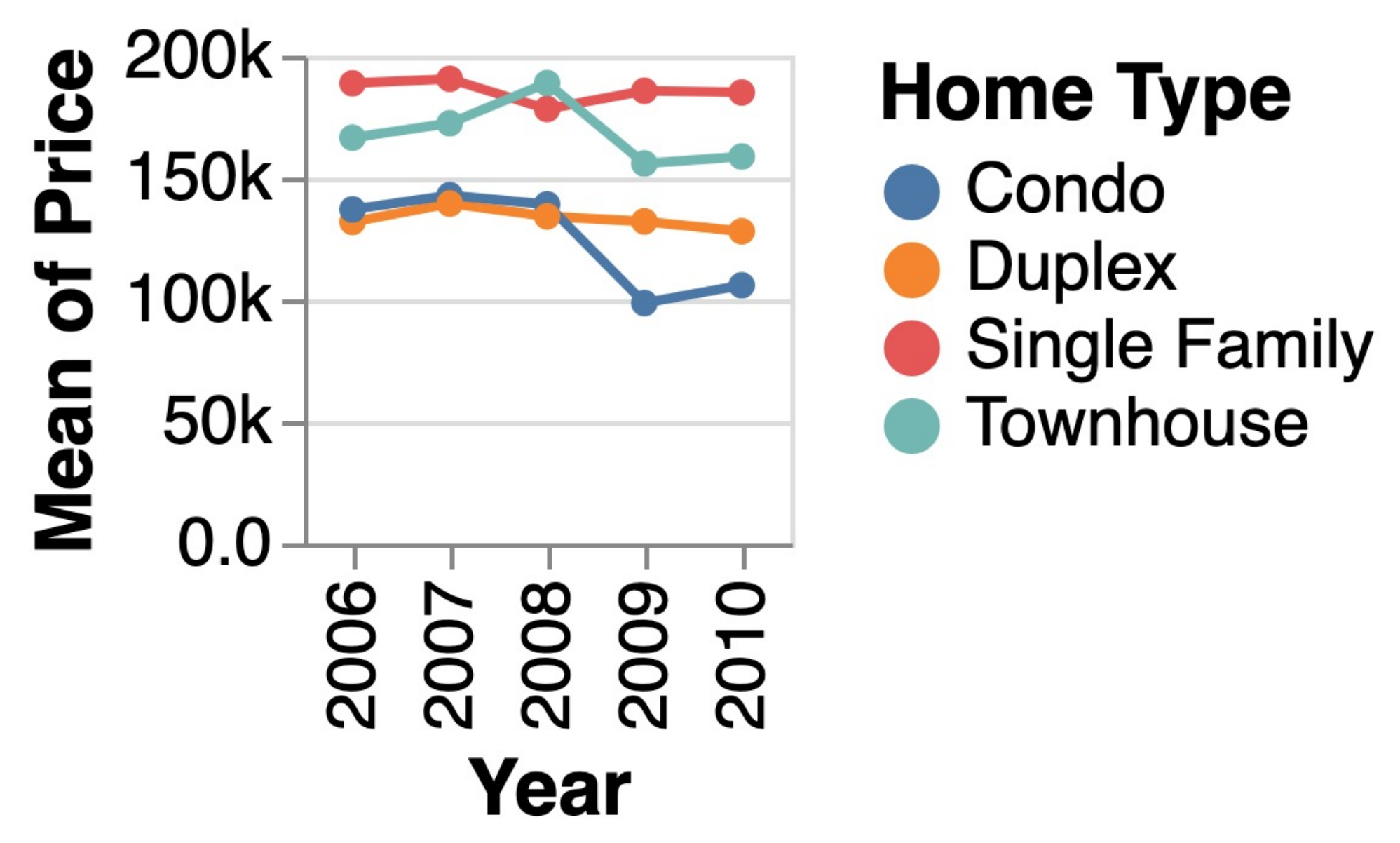}
\end{minipage}

\vspace{-1em}

\begin{minted}
    [
    baselinestretch=1,
    fontsize=\small,
    xleftmargin=15pt,
    linenos,
    breaklines,
    escapeinside=||,
    % ,style=tango
    ]
    {python}
# this query is automatically inferred as a follow-up.|\setcounter{FancyVerbLine}{6}|
resp_2 = nl4dv_instance.analyze_query("As a bar chart.", dialog="auto")
print(resp_2)
\end{minted}

\vspace{-1em}

\begin{minipage}[!b]{4cm}
    \begin{minted}[fontsize=\small,style=tango]{json}
    {
        "dialogId": "0",
        "queryId": "1", 
        "followUpConfidence": 
          "high", ...
    }
    \end{minted}
\end{minipage}\begin{minipage}[!b]{4.7cm}
    \centering
    \includegraphics[width=4.7cm]{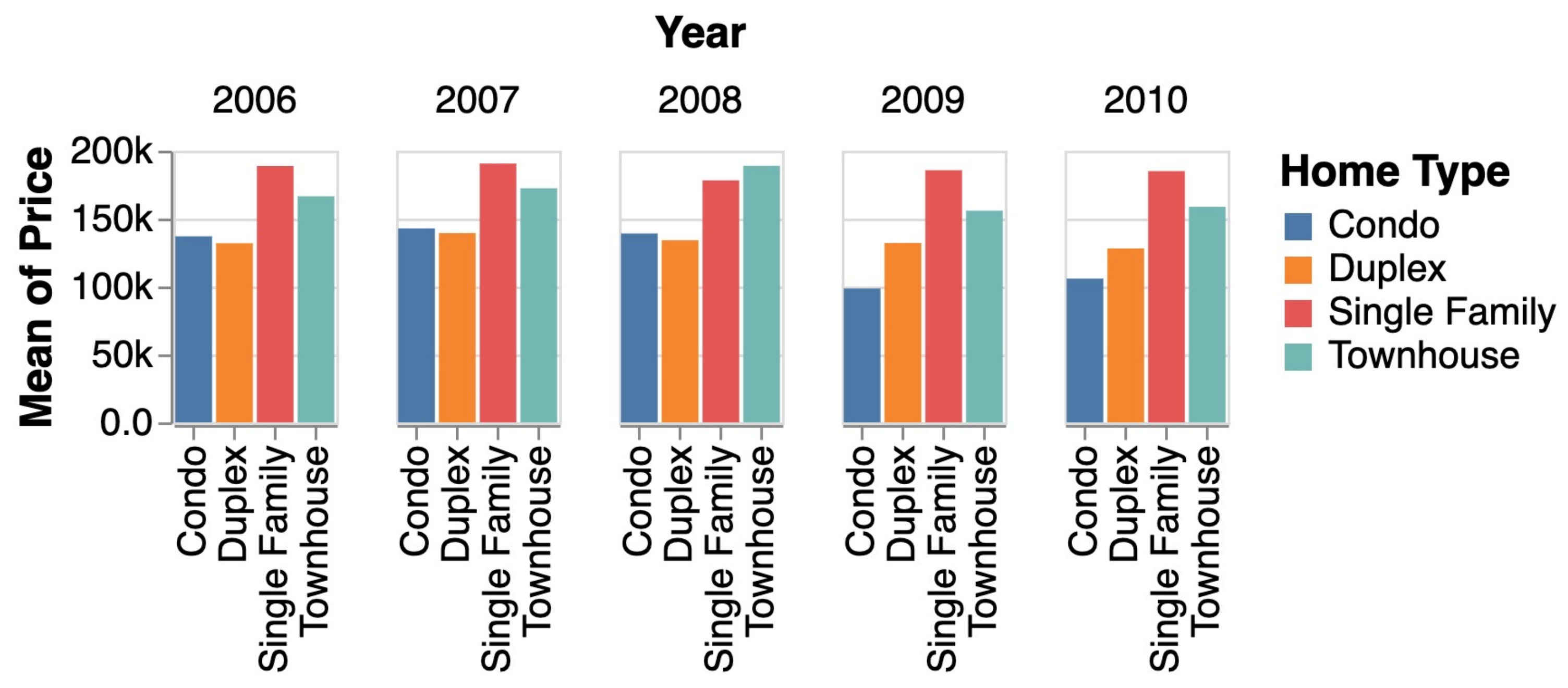}
\end{minipage}

\vspace{-1em}

\begin{minted}
    [
    baselinestretch=1,
    fontsize=\small,
    xleftmargin=15pt,
    linenos,
    breaklines,
    escapeinside=||,
    % ,style=tango
    ]
    {python}
# this query is a new, standalone query.|\setcounter{FancyVerbLine}{9}|
resp_3 = nl4dv_instance.analyze_query("Correlate Price and Lot Area.", dialog=False)
print(resp_3)
\end{minted}

\vspace{-1em}

\begin{minipage}[!b]{3.7cm}
    \begin{minted}[fontsize=\small,style=tango]{json}
    {
        "dialogId": "1",
        "queryId": "0", 
        ...
    }
    \end{minted}
\end{minipage}\begin{minipage}[!b]{4.1cm}
    \centering
    \includegraphics[width=3.5cm]{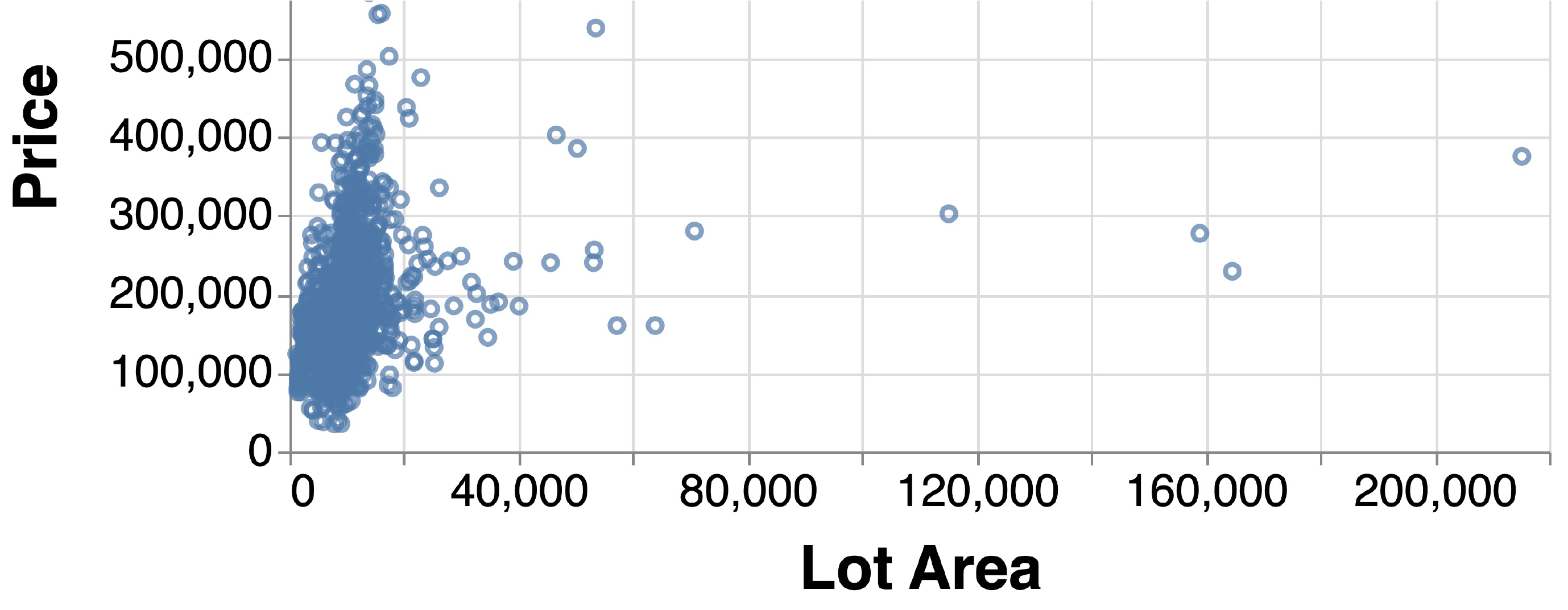}
\end{minipage}

\vspace{-1em}

\begin{minted}
    [
    baselinestretch=1,
    fontsize=\small,
    xleftmargin=15pt,
    linenos,
    breaklines,
    escapeinside=||,
    % ,style=tango
    ]
    {python}
# this query follows up a specific, older query.|\setcounter{FancyVerbLine}{12}|
resp_4 = nl4dv_instance.analyze_query("Just show condos and duplexes.", dialog=True, dialog_id="0", query_id="1")
print(resp_4)
\end{minted}

\vspace{-1em}

\begin{minipage}[!b]{4cm}
    \begin{minted}[fontsize=\small,style=tango]{json}
    {
        "dialogId": "0",
        "queryId": "2", 
        ...
    }
    \end{minted}
\end{minipage}\begin{minipage}[!b]{4.1cm}
    \centering
    \includegraphics[width=4.1cm]{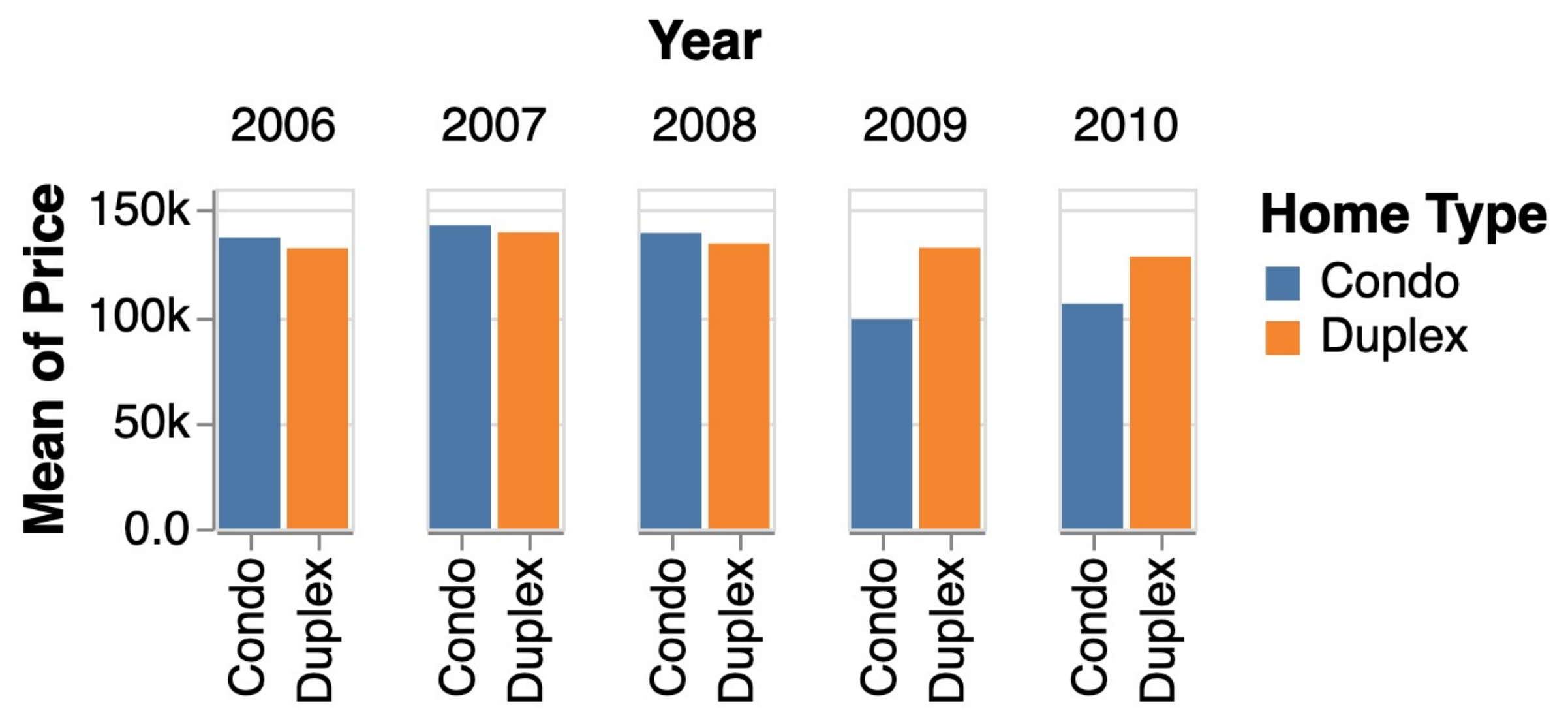}
\end{minipage}

\vspace{-0.5em}

\caption{\add{Python code illustrating how developers can enable conversational interaction in their applications using \app.} \vspace{-1em}}
\label{listing:init-eg}
\end{listing}

\begin{figure}
    \centering
    \includegraphics[width=\linewidth]{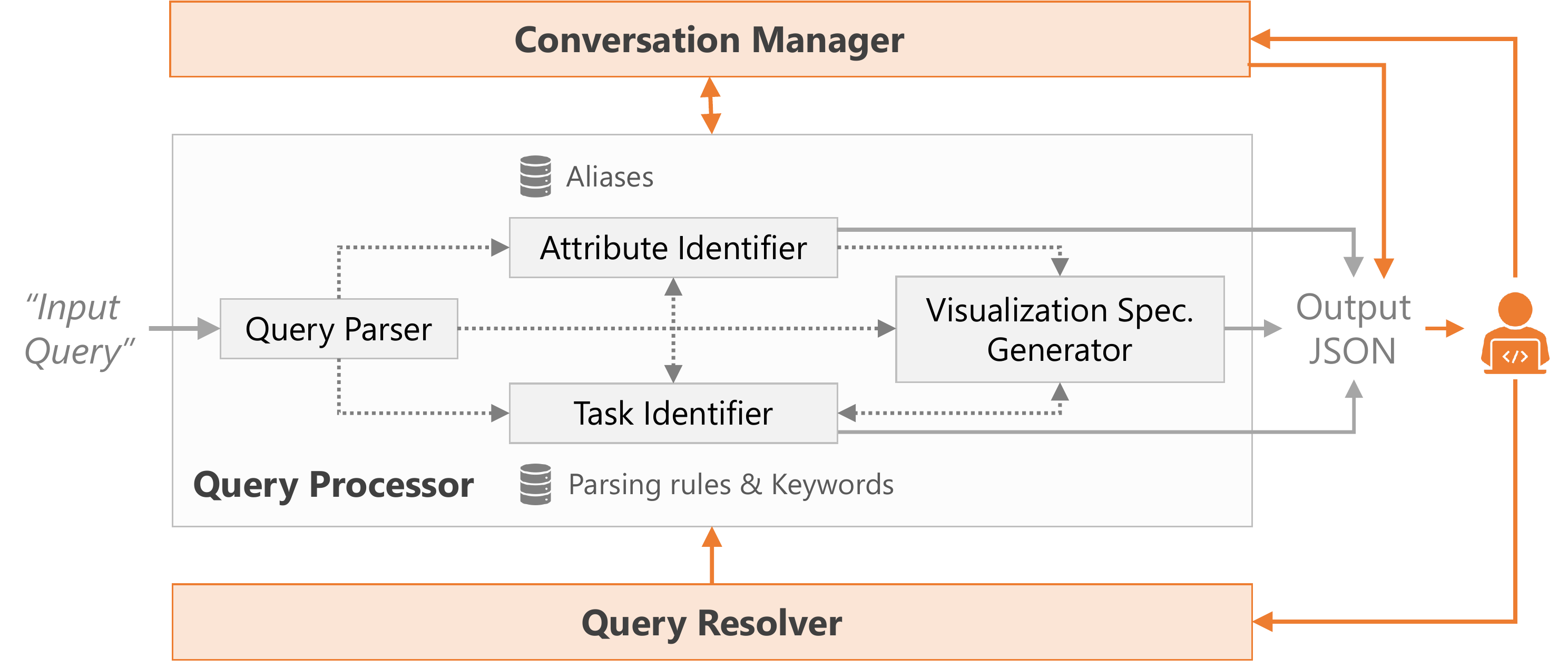}
    \vspace{-1.25em}
    \caption{Original \app architecture~\cite{narechania2020nl4dv} (in \textcolor{gray}{gray}) extended to support conversational interaction (in \textcolor{orange}{orange}). The arrows indicate the flow of information between the modules.
    }
    \label{fig:architecture}
\end{figure}


Listing~\ref{listing:init-eg} illustrates the basic Python code for \add{developers to enable} conversational interaction \add{in their own applications} using \app. Given a tabular dataset on \add{Houses (adapted from~\cite{de2011ames}; accessible at~\cite{housingdataset})} and a query string \add{specified by the end-user}, \inputParamValue{\emph{``Show average prices for different home types over the years''}}, with a single function call \function{analyze\_query(query)} (lines 1-3), \app first determines it as a standalone query (as it is the very first query), extracts data attributes and analytic tasks, recommends visualizations, and then assigns new \remove{variables}\add{objects} that identify that conversation (\variable{dialogId}$=$\val{``0''}) and the corresponding query (\variable{queryId}$=$\val{``0''}) as part of the output JSON (lines 4-5). 
After observing the output visualization,\remove{now wanting} \add{if the end-user wants a bar chart instead of a line chart, they may ask}, \inputParamValue{\emph{``As a bar chart''}} with a new parameter, \textbf{dialog}$=$\val{``auto''}. \app automatically determines this as a follow-up to the previous query (with a heuristically determined \variable{followUpConfidence}$=$\val{``high''}) and directly modifies its analytic specification, retaining the \variable{dialogId}$=$\val{``0''} but generating a new, now incremented \variable{queryId}$=$\val{``1''} as the second query in the conversation (lines 6-8). 
\add{If the end-user is} suddenly curious about how house prices compare with area, \add{they may ask}, \inputParamValue{\emph{``Correlate price and area''}}, explicitly specifying the query as standalone (\textbf{dialog}$=$\textcolor{inputParamColorBool}{\textbf{False}}). This time, \app increments \variable{dialogId}$=$\val{``1''} and resets \variable{queryId}$=$\val{``0''} since this is now the first query of a new, second conversation (lines 9-11). 
If the \add{end-user wants to resume their original conversation and only focus on certain home types, they may ask}, \inputParamValue{\emph{``Just show condos and duplexes''}}, this time explicitly specifying the query as a follow-up (\textbf{dialog}$=$\textcolor{inputParamColorBool}{\textbf{True}}) 
with additional parameters: \textbf{dialog\_id}$=$\inputParamValue{``0''}, \textbf{query\_id}$=$\inputParamValue{``1''}, that correspond to the first conversation (lines 12-14). As expected, the resultant \variable{dialogId}$=$\val{``0''} and \variable{queryId}$=$\val{``2''}, along with the filtered bar chart.

To achieve this kind of conversational interaction, we extended \app~\cite{narechania2020nl4dv}; 
\add{Figure~\ref{fig:architecture} illustrates the modified technical architecture.
The existing \textbf{Query Processor} module parses the input NL query using NLP techniques such as tokenizing and parts of speech tagging (\emph{Query Parser}), extracts data attributes through semantic and syntactic similarity matching (\emph{Attribute Identifier}) and analytic tasks through dependency parsing (\emph{Tasks Identifier}), and recommends relevant visualizations based on heuristics used in prior systems~\cite{mackinlay2007show, wongsuphasawat2015voyager, wongsuphasawat2017voyager}
(\emph{Visualization Specification Generator}), that are combined into an \emph{Output JSON}.} 
\add{The new \textbf{Conversation Manager} module enables developers to automatically determine or manually specify} a query as a follow-up (or not). This module also determines the type of follow-up (e.g., \emph{add} or \emph{remove} attributes), managing all operations on the internal\remove{ conversational} data structures. \add{Also new, the \textbf{Query Resolver} module} \remove{allows end-users to resolve}\add{facilitates resolving} NL ambiguities (e.g., by ``medals'' did the end-user mean ``\{Total $\vert$ Gold $\vert$ Silver $\vert$ Bronze\} Medals''?).

\begin{listing}[t!]

    \begin{minted}
        [
        baselinestretch=1,
        % fontsize=\scriptsize,
        fontsize=\small,
        xleftmargin=15pt,
        linenos,
        breaklines,
        escapeinside=||,
        % ,style=tango
        ]
        {python}
all_dialogs = {
 "0": [{"query":"show the distribution of salaries as a boxplot",...}, {"query":"How about goals instead?",..}],  
 "1": [{"query": "Show average goals per country",...}, {"query": "now group by foot",...}],
 "2": [{"query": "correlate age and salary",...}, {"query": "now show only defenders",...}],
 "2.0.0": [{"query": "correlate age and salary",...}, {"query": "what about only goalkeepers?",...}] # follow-up to query with dialog_id="2" and index=0
}
    \end{minted}
    \vspace{-1.5em}
    \caption{\textcolor{black}{Data structure to store multiple conversations, including branches (multiple follow-ups to the same query).} \vspace{-1em}}
    \label{listing:datastructure}
\end{listing}

\subsection{Facilitating Multiple Simultaneous Conversations}
Following the dialog shown in Listing~\ref{listing:init-eg}, whenever the end-user asks a new, standalone query, the \textbf{dialog\_id} is also incremented by \val{``1''} and the \textbf{query\_id} is re-initialized to \val{``0''} (identifiers are stringified after incrementing for efficient handling of data),
creating a new dialog instance that is uniquely identifiable by \variable{dialogId} and \variable{queryId}.
Subsequently, developers can explicitly follow up on specific queries by passing the follow-up query string along with additional input parameters: \textbf{dialog} (a boolean flag expressing an explicit intent to follow-up), \textbf{dialog\_id}, and \textbf{query\_id} to \function{analyze\_query(query)}. 

This design also enables end-users to ask multiple unrelated follow-ups to the same query. To create such conversational branches, developers can provide the same \textbf{dialog\_id} and \textbf{query\_id} in repeated calls to \function{analyze\_query(query)}. Internally, \app creates the desired branch point and outputs a new, unique \variable{dialogId} with the format: \val{``\{dialog\_id\}.\{query\_id\}.\{branch\_id\}''} (similar to the semantic versioning format~\cite{semver}), where \val{\{branch\_id\}} is the index of the branch stemming from the input parameters: \val{\{dialog\_id\}} and \val{\{query\_id\}}. This naming convention effectively represents the hierarchy of all entities involved in the conversation. Listing~\ref{listing:datastructure} shows how \app stores these conversations in a\remove{tree-like data structure, implemented as a} Python dictionary of lists with \textbf{dialog\_id}s as the keys and \textbf{query\_id}s as the indexes of the corresponding list of queries. This data structure enables efficient retrieve, append, modify, and delete operations. Note that calling \function{analyze\_query(query, dialog$=$\textcolor{inputParamColorBool}{\textbf{True}})}, without \textbf{dialog\_id} or \textbf{query\_id}, will make \app follow up on the \emph{most recent} \variable{dialogId} and \variable{queryId}; if these too do not exist (e.g., it is the very first conversation), then an error is thrown.

\subsection{Detecting, Classifying and Processing Follow-ups}
To supply \textbf{dialog}, \textbf{dialog\_id} and \textbf{query\_id} parameters to \function{analyze\_query(query)}, developers have to provide GUI affordances for end-users, e.g., a checkbox to specify if \textbf{dialog}$=$\val{True} or not (and which conversation to follow-up on), which can be an unnatural end-user experience. 
To alleviate this, \app offers a \textbf{dialog}$=$\val{``auto''} setting (overloading the otherwise boolean input data type) that automatically determines if the query is a follow-up or not and outputs a \variable{followUpConfidence} rating: \{\val{``high''}, \val{``low''}, \val{``none''}\} reflecting \app's confidence in making the inference. 
This rating is heuristically determined based on the previous query, an \textbf{explicit\_followup\_keywords} map -- keywords that convey natural conversational intents to follow-up (e.g., ``add'', ``replace''), and an \textbf{implicit\_followup\_keywords} map -- keywords that implicitly convey an intent to follow-up (e.g., ``instead of'', ``only'').
\add{The \textbf{implicit\_followup\_keywords} are further classified as \emph{non-ambiguous} -- keywords that always convey an intent to follow-up (e.g. ``instead of'', ``rather than'') and \emph{ambiguous} -- keywords that can occur in a follow-up as well as standalone context (e.g., ``only'').
\app assigns queries containing \textbf{explicit} keywords or \textbf{implicit} \emph{non-ambiguous} keywords with a \textbf{high} \variable{followUpConfidence} rating and \textbf{implicit} \emph{ambiguous} keywords with a \textbf{low} \variable{followUpConfidence} rating.
For queries with no matching keywords, \app compares the attributes, tasks, and visualizations of the current and the previous query and based on a heuristics and rule-based decision tree, assigns either a \textbf{low} or \textbf{none} \variable{followUpConfidence} rating, the latter corresponding to a new, standalone query. For example,
a query \inputParamValue{\emph{``Show the average now.''}} is a compatible follow-up to its predecessor, \inputParamValue{\emph{``Show maximum price across different home types.''}}; the desired change from ``maximum'' to ``average'' in the absence of any other follow-up keywords or attributes makes them compatible.}
Developers can override these default maps by supplying custom \textbf{explicit\_followup\_keywords} and \textbf{implicit\_followup\_keywords} objects 
through the \function{NL4DV()} constructor during initialization.

Next, the \textbf{explicit\_followup\_keywords} map classifies the follow-up query as one of three types: \emph{add}, \emph{remove}, or \emph{replace} (inspired by Evizeon's \emph{continue}, \emph{retain}, \emph{shift} transitional states~\cite{hoque2017applying}) and maps it to one or more components of an analytic specification: \emph{data attributes}, \emph{analytic tasks}, and \emph{visualizations}.
Note that the resultant combinations (e.g., \emph{replace} + \emph{data attribute}) are not always mutually exclusive, e.g., replacing an attribute can sometimes also modify the task(s) and/or the visualization(s). Lastly, \app references the parent query's (the query being followed upon) analytic specification and makes necessary associations (e.g., creating new conversational branches) and modifications (e.g., dropping an existing attribute), eventually generating a new specification as a JSON object.

By configuring the keyword maps and supplying methods with appropriate parameters, end-users can \emph{add}, \emph{remove}, or \emph{replace} data attributes, either \textbf{explicit}ly, e.g., \emph{``Replace budget with gross''}--which makes a direct reference to the data attributes and the follow-up task; or \textbf{implicit}ly, e.g., \emph{``Now show only budget''}--which indirectly suggests to remove all other attributes except ``Production Budget''.
Unlike attributes, following up on \remove{analytic }tasks is different because end-users are unaware of the associated technical jargon, e.g. \emph{``Add Find Extremum to Worldwide Gross''} is not a natural query an end-user would ask; they would rather say, \emph{``Show me the highest grossing movie''}, which would then infer the \emph{Find Extremum} task~\cite{amar2005low} (through `highest'). Thus, most queries that follow-up on tasks are \textbf{implicit} in nature. \app currently supports \emph{sort} (e.g., \emph{``Sort by budget in an ascending order''}), \emph{find extremum} (e.g., \emph{``Which of these genres has the smallest budget?''}), \emph{filter} (e.g., \emph{``Now show only action movies''}), and \emph{derived value} (e.g., \emph{``Replace average with sum''}) tasks~\cite{amar2005low}. 
A follow-up to \emph{add} (or \emph{remove}) a visualization is meaningless as there will (or must) always be some recommended chart. 
\emph{Replace} is the only meaningful task and it can be \textbf{explicit} (e.g., \emph{``Replace this line chart with a bar chart''}) or \textbf{implicit} (e.g., \emph{``As a bar chart instead''}).

\subsection{Resolving Ambiguities during Query Interpretation}
\begin{listing}[t!]

    \begin{minted}
        [
        baselinestretch=1,
        fontsize=\small,
        linenos,
        xleftmargin=15pt,
        breaklines,
        escapeinside=||,
        % ,style=tango
        ]
        {python}
from nl4dv import NL4DV
nl4dv_instance = NL4DV(data_url="olympic_medals.csv")
init_response = nl4dv_instance.analyze_query("Show medals in hockey and skating by country.")
# Multiple ambiguities are detected from the query.
print(init_response)
\end{minted}
    
    \vspace{-2em}
    
    \begin{minted}[breaklines,fontsize=\small,style=tango]{json}
    { "dialogId": "0", "queryId": "0",
      "ambiguities": {
        "attribute": {
          "medals": { "options": ["Bronze Medal","Gold Medal","Silver Medal", "Total Medal"], 
                      "selected": null}},
        "value": {
          "skating": { "options": ["Figure Skating", "Short Speed Skating", "Speed Skating"], 
                       "selected": null},
          "hockey": { "options": ["Hockey", "Ice Hockey"], 
                      "selected": null}}
      }, ... }
    \end{minted}
    
    \vspace{-2em}
    
    \begin{minted}
        [
        baselinestretch=1,
        fontsize=\small,
        linenos,
        xleftmargin=15pt,
        breaklines,
        escapeinside=||,
        % ,style=tango
        ]
        {python}
resolved_response = nl4dv_instance.update_query({"attribute": { "medals": "Total Medal" }, "value": {"skating": "Speed Skating", "hockey": "Ice Hockey"}})|\setcounter{FancyVerbLine}{6}|
print(resolved_response)
# The "selected" property is updated in the response.
    \end{minted}
    
    \vspace{-2em}
    
    \begin{minted}[breaklines,fontsize=\small,style=tango]{json}
    {"dialogId":"0","queryId":"0","ambiguities":{...}, ...}
    \end{minted}
    
    \vspace{-1.5em}
    
    \caption{\textcolor{black}{Python code illustrating how \app helps resolve ambiguities via \function{update\_query(obj)} (line 6). 
    } \vspace{-1em}}
    \label{listing:ambiguity-resolution-eg}
    \end{listing}
Natural language (NL) is often ambiguous and underspecified, e.g., consider the query,
\emph{``Show medals in hockey and skating by country''} regarding a dataset on Olympic Medals \add{(adapted from~\cite{kaggleolympicsdataset}; accessible at~\cite{olympicsdataset})}. Here, ``medals'' (\textbf{attribute}) could be mapped to either of [``Total Medals'', ``Gold Medals'', ``Silver Medals'', ``Bronze Medals''], ``hockey'' (\textbf{value}) could be mapped to either of [``Ice Hockey'', ``Hockey''], and ``skating'' could be mapped to either of [``Figure Skating'', ``Speed Skating'', ``Short Speed Skating'']. 

These ambiguities can cause problems 
while processing ambiguous follow-up queries (e.g., \emph{``Sort by medals''}--Which type of ``medals''?), and hence must be resolved a priori. \app detects these \textbf{attribute}-level and \textbf{value}-level ambiguities and makes them accessible in the output JSON under a new key, \variable{ambiguities}. In addition, \app now provides a new function \function{update\_query(obj)}, to help developers design experiences that resolve ambiguities directly through the toolkit, also enabling accurate processing of subsequent follow-up queries. Listing~\ref{listing:ambiguity-resolution-eg} illustrates how \function{update\_query(obj)} (line 6) takes a Python dictionary as input, that includes the types of ambiguities (\inputParamValue{``attribute''} and \inputParamValue{``value''}), the corresponding keywords in the query (\inputParamValue{``medals''}, \inputParamValue{``hockey''}, and \inputParamValue{``skating''}), and the corresponding entities selected by the end-user for resolution.
\app then updates the \variable{selected} entities under \variable{ambiguities} as well as the \textbf{attributeMap} and the \textbf{taskMap}, recommending a new \textbf{visList} (visualizations). Note that developers may not always provide end-users with affordances to resolve such ambiguities. In these cases, \app automatically resolves ambiguities by itself, \remove{internally calling \function{update\_query(obj)} with} \add{selecting the entities that have the highest string-based similarity score with the corresponding query keyword}, and calling \function{update\_query(obj)}. 
In case of ties, entities that were detected first are selected. 



\section{Creating Visualization Systems with \app}
\label{section:example-applications}
\subsection{NL-Driven Vega-Lite Learner}
\begin{figure}[t]
    \centering
    \includegraphics[width=\linewidth]{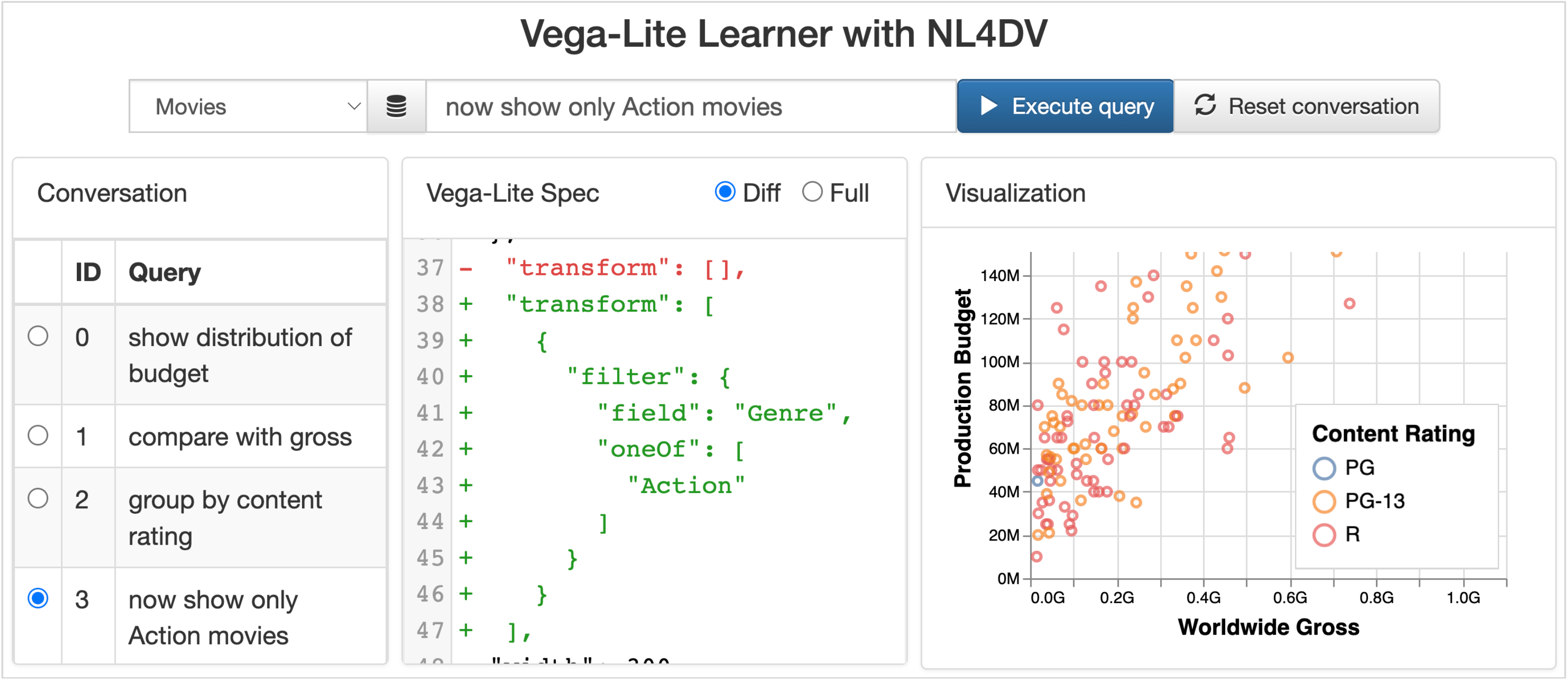}
    \vspace{-1.25em}
    \caption{An NLI that helps users learn Vega-Lite syntax (e.g., the \emph{transform} property to apply filters), through NL queries.}
    \label{fig:vl-learner}
\end{figure}

The NL-Driven Vega-Lite Editor in NL4DV~\cite{narechania2020nl4dv} demonstrated how NL can be used to create, edit, and hence learn the Vega-Lite~\cite{satyanarayan2016vega} grammar. However, end-users of this system need to be proficient with the Vega-Lite syntax (e.g., properties and operators) to be able to successfully edit the specifications output by \app.
Figure~\ref{fig:vl-learner} illustrates the user interface of a similar NL-Driven Vega-Lite Learner that demonstrates how conversational interaction can help users learn this grammar \emph{better} \add{by sequentially processing short, specific NL intents and incrementally building the Vega-Lite specification, helping users learn the syntax changes required to achieve the corresponding intents}. 
Users ask a series of short NL queries and observe the resultant Vega-Lite specifications.
These queries are chained together, forming a conversation. Users can see the \emph{diff} (i.e., added and removed entities) between the Vega-Lite specifications of the selected query and its predecessor through code highlights (green implies addition; red implies deletion). For example, a follow-up query to apply a filter, \emph{``Now show only Action movies''} 
generates a Vega-Lite specification that differs from the previous query's specification in terms of the \emph{transform} property, helping the user learn how Vega-Lite filters are specified.
To develop this interface, developers can sequentially call \function{analyze\_query(\textbf{query}, \textbf{dialog}$=$\textcolor{inputParamColorBool}{\textbf{True}})} and then focus on computing the \emph{diffs} between the Vega-Lite specifications of the query and its predecessor and
\add{programming the layout, styling, and interactivity aspects using HTML, CSS, JavaScript.}
\remove{designing the user interface using HTML, CSS, JavaScript: presenting the input queries, corresponding Vega-Lite specifications, their \emph{diff}s, and the rendered visualizations.}



\subsection{Mind Mapping Conversations about a Dataset}
\label{sec:mindmapping}
\begin{figure}[t]
    \centering
    \includegraphics[width=\linewidth]{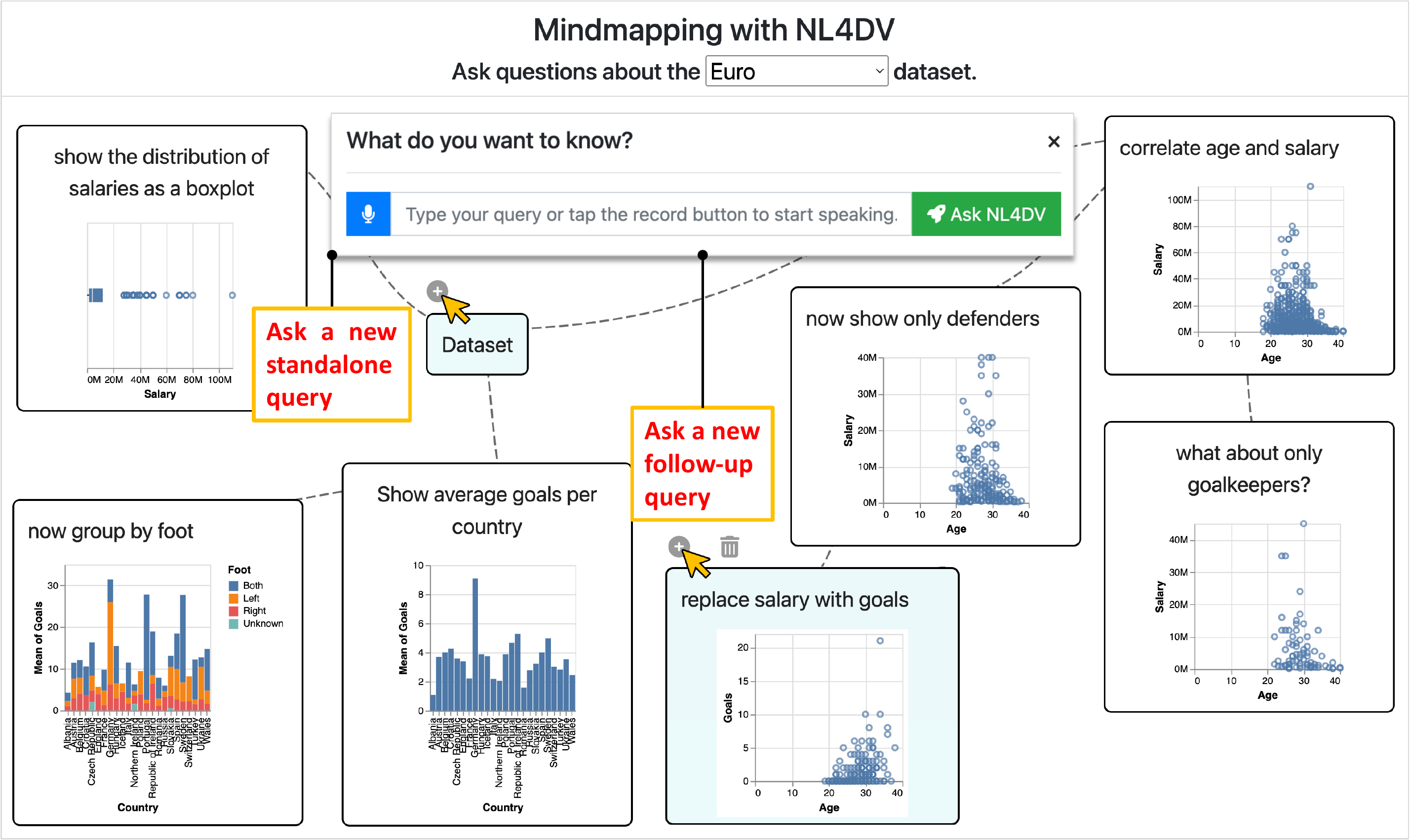}
    \vspace{-1.25em}
    \caption{A mind mapping app that enables users to have free-flowing conversations about a dataset. Three dialogs with follow-ups are connected to the ``Dataset'' via dashed lines.}
    \label{fig:mindmap}
\end{figure}

In this second use-case, we demonstrate how the input parameters: \textbf{dialog}, \textbf{dialog\_id} and \textbf{query\_id} in \function{analyze\_query(query)} can help end-users engage in multiple simultaneous conversations, unlike the NL-Driven Vega-Lite Learner, that supports only one conversation at-a-time.
Figure~\ref{fig:mindmap} illustrates the user interface of a mind mapping application that helps users engage in free-flowing conversations regarding a European soccer players dataset \add{(adapted from~\cite{nytimes2014wc}; accessible at~\cite{eurodataset})}. Listing~\ref{listing:datastructure} shows the corresponding data structure maintained by \app. Through speech or text input, users can ask standalone queries (e.g., \emph{``Correlate age and salary''}) as well as follow-up queries (e.g., \emph{``Now show only defenders''}) by clicking the plus icon, enabled by hovering on the corresponding mind map node (the rectangular block). Users can also follow-up on already followed-up queries, forming new conversational branches (e.g., \emph{``What about only goalkeepers?''}). 
To develop this interface, developers can call \function{analyze\_query(\textbf{query}, \textbf{dialog}, \textbf{dialog\_id}, \textbf{query\_id})}, supplying the dialog and query identifiers based on the corresponding query that is to be followed-upon.
Then, based on the newly generated \variable{dialogId} and \variable{queryId}, a new node is created and appended to the corresponding predecessor query node. 


\subsection{Collaboratively Resolve Ambiguities in a ChatBot}
\begin{figure}[t]
    \centering
    \includegraphics[width=\columnwidth]{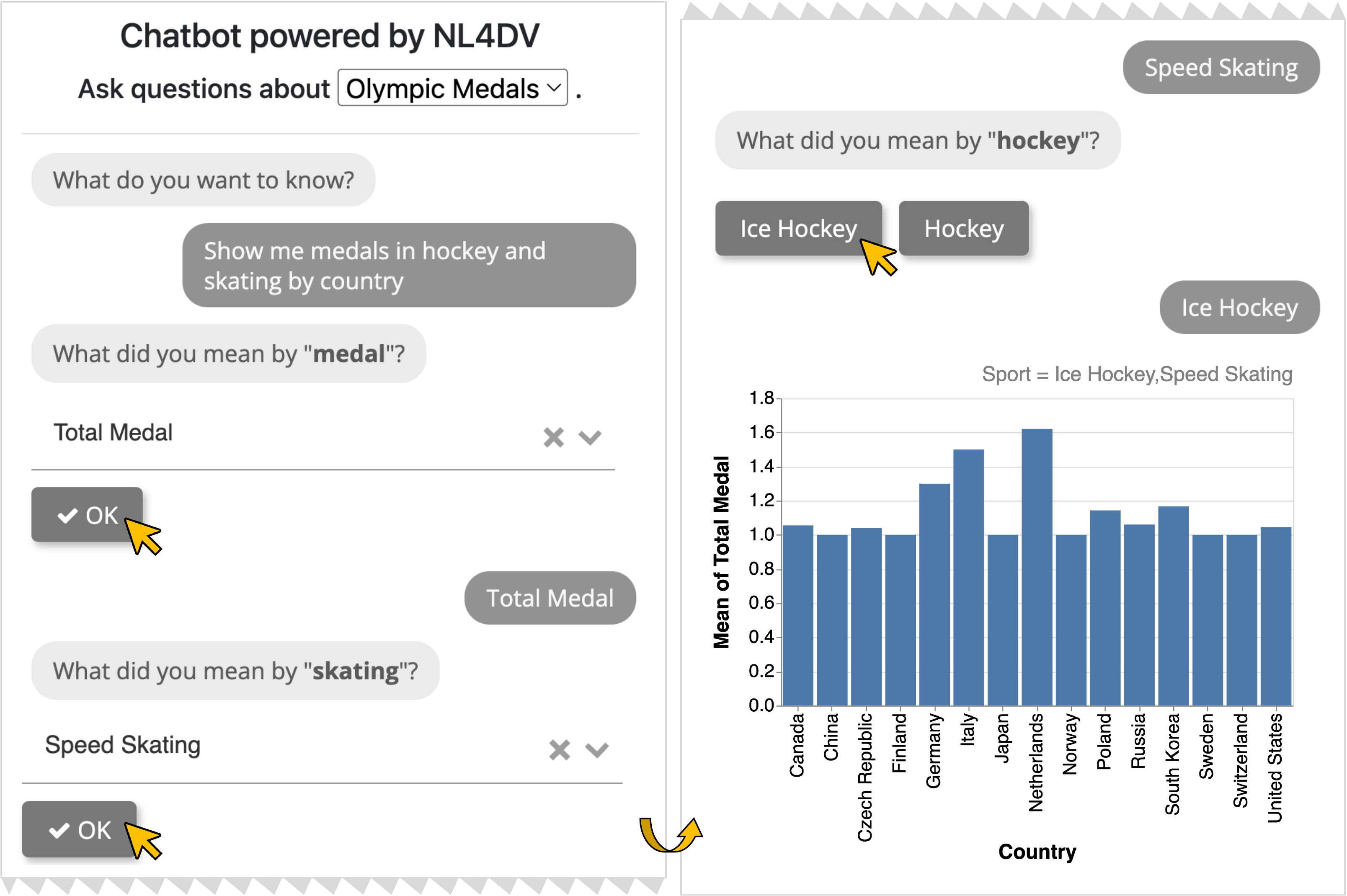}
    \vspace{-1.25em}
    \caption{A chatbot where the system collaborates with the user to resolve ambiguities during query interpretation.}
    \label{fig:chatbot}
\end{figure}


In this third use-case, we demonstrate how \app's \textbf{Query Resolver} can help resolve ambiguities that often occur in natural language. 
Figure~\ref{fig:chatbot} illustrates a standard chatbot user interface that presents DataTone-like~\cite{gao2015datatone} ``ambiguity widgets''--dropdowns and buttons. End-users can disambiguate by interacting with the widgets, notifying \app through a function call to \function{update\_query(obj)}. After all ambiguities are resolved, the system renders the now-unambiguous visualization. 
To develop this interface, developers can loop through the \variable{ambiguities} object in the output JSON and present corresponding \variable{options} to the end-user, e.g., as options in a select dropdown. As the end-user makes their choices, a function call to \function{update\_query(obj)} will resolve the ambiguity, updating the \variable{selected} properties in the output JSON. Listing~\ref{listing:ambiguity-resolution-eg} illustrates this data exchange between the user interface and \app.


\section{Conclusion, Limitations, and Future Work}
\label{section:discussion}

In this work, we extend an existing natural language (NL) for data visualization toolkit, \app, to enable developers to integrate conversational interaction capabilities within natural language interfaces. 
We demonstrate \app's capabilities through three examples and open-source the toolkit at \textbf{\url{https://nl4dv.github.io/nl4dv/}}.

While testing, we noted certain conversational ambiguities, e.g., if a query, \emph{``Show only Action movies''} is followed-up with \emph{``What about R-rated movies?''} does the user mean to augment the previous filter or replace it with the new one? 
Consider another query, \emph{``Visualize budget distribution as a histogram instead of a boxplot''}; here, the user means to ask a standalone query, but the presence of ``instead of'' (an implicit follow-up keyword) will make \app wrongly treat it as a follow-up. 
We will address these ambiguities and translation errors in future releases.
We also plan a formal performance evaluation of the toolkit. However, unlike conversational text-to-SQL dataset benchmarks (e.g., CoSQL~\cite{yu2019cosql}), there are currently no such benchmarks for visualization tasks. An area of future work, thus, for current text-to-visualization datasets~\cite{fu2020quda,srinivasan2021collecting,luo2021synthesizing}, that focus on singleton utterances, is to include multi-turn utterances.


\label{section:conclusion}


\acknowledgments{\add{This work was supported in part by an NSF Grant IIS-1717111. We thank Arjun Srinivasan and the Georgia Tech Visualization Lab.}}

\bibliographystyle{abbrv-doi}

\bibliography{template}
\end{document}